\documentclass[%
 aip,
 jmp,
 amsmath,amssymb,
 reprint
]{revtex4-1}

\usepackage{graphicx}
\usepackage{dcolumn}
\usepackage{bm}
\usepackage{color}
\usepackage{epstopdf}
\usepackage{amsmath}
\usepackage{mathtools,ulem}

\usepackage{natbib}

\begin{document}

\preprint{AIP/123-QED}

\title[Parametric study and optimization trends]{Parametric study and optimization trends for the Von-K\'{a}rm\'{a}n-Sodium dynamo experiment}

\author{J. Varela}
\email{rodriguezjv@ornl.gov}
\affiliation{Oak Ridge National Laboratory, Oak Ridge, Tennessee 37831-8071}
\affiliation{M\'ecanique et les Sciences de l'Ing\'enieur, LIMSI, CNRS, Univ. Paris-Sud, Universit\'e Paris-Saclay, B\^at 508, Campus Universitaire F-­91405 Orsay}
\affiliation{AIM, CEA/CNRS/University of Paris 7, CEA-Saclay, 91191 Gif-sur-Yvette, France}

\date{\today}

\begin{abstract}
We present magneto-hydrodynamic simulations of liquid sodium flow with the PLUTO compressible MHD code. We investigate the influence of the remnant magnetic field orientation and intensity, impinging velocity field due to Ekman pumping as well as the impeller dimensions on the magnetic field collimation by helical flows in between the impeller blades. For a simplified cartesian geometry we model the flow dynamics of a multi-blades impeller inspired by the Von-K\'{a}rm\'{a}n-Sodium (VKS) experiment. The study shows that a remnant magnetic field oriented in the toroidal direction is the less efficient configuration to collimate the magnetic field, although if the radial or vertical components are not negligible the collimation is significantly improved. If the intensity of the remnant magnetic field increases the system magnetic energy is larger but the magnetic field collimation efficiency is the same, so the gain of magnetic energy is smaller as the remnant magnetic field intensity increases. The magnetic field collimation is modified if the impinging velocity field changes: the collimation is weaker if the impinging velocity increases from $\Gamma=0.8$ to $0.9$, slightly larger if the impinging velocity decreases from $\Gamma=0.8$ to $0.7$. The analysis of the impeller dimensions points out that the most efficient configuration to collimate the magnetic field requires a ratio between impeller blade height and base longitude between $0.375$ - $0.5$. The largest enhancement of the hypothetical $\alpha^2$ dynamo loop, compared to the hypothetical $\Omega$-$\alpha$ dynamo loop, is observed for the model that mimics TM 73 impeller configuration rotating in the unscooping direction with a remnant  magnetic field of $10^{-3}$ T orientated in the radial or vertical direction. The optimization trends obtained in the parametric analysis are also confirmed in simulations with higher resolution and turbulence degree.
\end{abstract}


\maketitle

\section{Introduction \label{sec:introduction}}

The dynamo effect consists in the transformation of mechanical energy to magnetic energy~\citep{Moffatt78,2004ApJ...614.1073B}, the main source of magnetic fields in the nature \citep{2006GApFD.100..281N,ZAMM:ZAMM19840640913}. The dynamo action takes place in celestial bodies such as the Sun~\citep{Ossendrijver2003}, galaxies~\citep{Beck1996} and the Earth~\citep{Aubert2015}. In the dynamo action, toroidal and poloidal components of the magnetic field are coupled via complex conductive fluid flows by the dynamo loop, that consists in the regeneration of the toroidal field from the poloidal field and vice-versa.

The dynamo field can be analyzed in the framework of the mean field dynamo theory \citep{1980AN....301..101R,ZAMM:ZAMM19840640913}, based on the classical $\alpha$ effect driven by helicoidal motions (helical turbulence) and the $\Omega$ effect linked to the differential rotation (large scale shear flows) of the system. If the large scale shear flows are weak, for example in full convective stars such as spectral class M stars with a mass smaller than $0.35$ times the mass of the Sun and anti-solar differential rotation (faster rotation in the poles respect to the equator), recent numerical simulations show the generation of $\alpha^2$ dynamos in the buoyancy-dominated regime because the helical turbulence is dominant so the poloidal and toroidal magnetic field components are regenerated via the $\alpha$ effect \citep{Simitev,Yadav}. On the other side, if there is a balance between turbulent and meridional flows, the $\Omega$ effect mainly regenerates the magnetic field toroidal component andan $\alpha - \Omega$ dynamo is observed \citep{Brown,Brown2}. If the turbulent flows are dominant although the differential rotation of the system is non negligible, both $\alpha$ and $\Omega$ effects are important in the toroidal magnetic field regeneration so an $\alpha^{2} - \Omega$ dynamo is triggered \citep{Schubert,Chabrier,Dormy}.

The dynamo regime can be identified by the magnetic Prandtl number $P_{m}$, the ratio of kinetic viscosity and magnetic diffusivity. If $P_{m} \ll  1$, we describe the dense plasma of the Sun or liquid metal experiments where the magnetic field is large enough to affect the fluid motions \citep{2004ApJ...614.1073B}. On the other hand, if $P_{m} \gg 1$ we study a hot diffuse plasma in the interstellar medium or galaxy clusters, where the field enhancement is caused by the stretching of the magnetic field by the fluid motion \citep{Ott}.

A common way to study the dynamo action regime of important astrophysical or geophysical systems is with numerical models \citep{1.1331315,2000PhRvL..84.4365G,2007PhRvL..98d4502M}, although the applicability of such models is limited and must be complemented by experiment, for example liquid metal experiments. Among them, the Von-K\'{a}rm\'{a}n-Sodium (VKS) experiment consists in a cylindrical container of diameter $0.578$ m and length $0.604$ m filled with liquid sodium, where two 300 kW engines power up two co- or counter rotating co-axial propellers driving the liquid metal flows. The VKS experiment level of turbulence is comparable to celestial bodies, so "turbulence" could be the dominant mechanism for the generation of magnetic fields in both cases \citep{2009EL.....8739002G,PhysRevLett.101.104501}. Such a mechanism is driven by the interaction of the differential rotation and the non-axisymmetric velocity perturbations \citep{03091920701523410}, or the self-interaction of the helical perturbations\citep{PhysRevLett.109.024503}. 

The source of turbulence in the VKS experiment is the counter-rotation of two impellers ({\it{i.e.}} disks fitted with blades). Above a critical rotation frequency a dipolar magnetic field aligned with the symmetry axis of the set-up is spontaneously generated. It should be noted that the magnetic field is an axial dipole in average due to the presence of non-axisymmetric components driven by the flow turbulent fluctuations. Previous studies have shown that the dynamo threshold is at least 2 to 3 times smaller if the blades and/or the disks are made of copper, with large electrical conductivity, or with stainless steel \citep{PhysRevE.88.013002}, pointing out the essential effect of the impeller material in the dynamo mechanism \citep{PhysRevLett.101.104501,PhysRevLett.104.044503,PhysRevE.91.013008,Gissinger}. In addition, other studies revealed that if the curvature of the blades increases or the blade is made of magnetized ferromagnetic material the dynamo threshold decreases \citep{2014NJPh...16h3001F}. In previous communications the authors demonstrated that models with perfect ferromagnetic boundary conditions lead to a larger collimation of the magnetic field and a higher magnetic energy content of the system. These trends were confirmed in models with enhanced turbulence level in the fluctuating regime, therefore ferromagnetic configurations are the most interesting from the point of view of the experiment optimization \citep{PhysRevE.92.063015,Varela2017}. 

The present communication is dedicated to study the magnetic field collimation by helical flows in between the impeller blades for a model inspired in VKS experiment. We analyze the dependency of the magnetic field collimation with the orientation and intensity of the remnant magnetic field (RMF), different impinging velocity fields due to Ekman pumping, as well as the impeller base and blades dimensions. The aim of the analysis is to identify the parameter space that leads to the most efficient magnetic field collimation and the largest generation of magnetic fields, required to optimize VKS experiment operation. We use Magneto-HydroDynamic (MHD) numerical simulations in a simplified geometry mimicking the flow structure in the vicinity of a ferromagnetic impeller.

\section{Numerical model \label{sec:model}}

We use the PLUTO code with a resistive and viscous MHD single fluid model in 3D Cartesian coordinates \citep{2007ApJS..170..228M}. The VKS experiment geometry and the simulation domain are plotted in Figure 1. We simulate the helical flows near the impeller region in between two blades, with X, Y and Z directions corresponding to local azimuthal (toroidal), radial and vertical (poloidal) directions. For simplicity, we consider straight blades instead of curved blades and walls without thickness. The orange surfaces on Figure 1A and B represent the blades (at X$=0$ and X$=2$), the brown surface the impeller disk (at Z$=0$) and the gray surface the cylinder outer wall (at Y$=4$). In the models with different impeller base size "$D$", X value can be $1$, $1.5$, $2.0$ or $2.5$. Blade's geometry is taken into account via the velocity boundary condition, through $\Gamma$, the ratio of the poloidal to toroidal mean velocity that varies from $0.9$ to $0.46$ as the blade's curvature is changed from $34^o$ (unscooping sense of rotation) to $-34^o$(scooping sense of rotation), see table I and figure 3 of [F. Ravelet, 2005]. To analyze the effect of the impinging velocity field due to Ekman pumping we performed simulations with $\Gamma = 0.7$, $0.8$ and $0.9$. In addition, in several models we changed the blade height "$L$" using $L = 0.75$, $1.0$ and $1.25$ m fixing the Ekman pumping.

We impose in the impeller base and blades perfect ferromagnetic boundary conditions ($\vec{B} \times \vec{n} = \vec{0}$, with $\vec{n}$ the surface unitary vector), null velocity and constant slope (Neumann boundary conditions) for the density ($\rho$) and pressure ($p$). We only consider perfect ferromagnetic boundary conditions because the aim of the analysis is to find the optimization trend of VKS experiment, and a ferromagnetic impeller is the configuration that shows the most efficient magnetic field collimation. For the wall at $Y = 4$ and at the other boundaries, the magnetic field is fixed to $10^{-3} T$ and oriented in the azimuthal $\vec{X}$ direction, mimicking an azimuthal disk magnetization observed in the VKS experiment \citep{1367-2630-14-1-013044}. The value of $10^{-3} T$ has been chosen to match the order of magnitude of the remnant magnetic field observed in the impeller, after a dynamo has been switched off. In several models the RMF is varied to half, 5 or 10 times this value, and the orientation can be in the radial $\vec{Y}$ and vertical $\vec{Z}$ direction, with the aim to analyze the role of the RMF orientation and intensity on the magnetic field collimation. The RMF orientation and intensity depends on the magnetic field generated by the dynamo. Within the planes Z$=2$ and X$=0$ (outside the blade), the velocity is fixed to $\vec{V}=(10, 0, -10\Gamma) $ m/s, mimicking the impinging velocity field due to Ekman pumping towards the impeller. Outflow velocity conditions are imposed in the plane X$=2$ (outside the blade) and in the plane Z$=0$ (outside the impeller base). Velocity is null on the impeller and the container wall. This is a simple model of the expected global flow driven by the impellers rotation. We do not consider any further feedback effect between the system global flow and the local setup. This simplified model serves as an idealized representation of the cavity in between the impeller blades of the VKS experiment. It has been chosen with the sole purpose to model high degree of turbulence in this cavity using high resolution, not easily accessible in global setups. The density is fixed to $931$ kg/m$^{3}$ in the left wall outside the blade ($X=0$) and has a constant slope in the rest. The pressure is calculated as $p = \rho c^{2}_{s}/\gamma $ with $\gamma = 5/3$ the specific heat ratio and $c_{s} = 250$ m/s the sound speed.The $c_{s}$ value is one order of magnitude smaller than the real sound speed in liquid sodium to keep a time step large enough for the simulation to remain tractable. The impact on the simulations is small and the largely incompressible nature of the liquid sodium flow is preserved (subsonic low Mach number flow or pseudo-incompressibility regime), because the effective Mach number $M=\|\vec{V}\|/c_s \approx 0.06$ is lower than the transition Mach number of $0.3$ between incompressible and subsonic flows.

The numbers of grid points are typically $128$ in the (X) and (Z) directions and $256$ in the (Y) direction for the simulations with kinetic Reynolds number $R_{e}=\rho V L/\nu =200$ ($L=1$ m and $\nu$ the dynamic viscosity). We also perform simulations of higher grid resolution, $256$ in the (X) and (Z) directions and $512$ in the (Y) direction with kinetic Reynolds number $R_{e}=1000$, to confirm the trends obtained in the parametric analysis for a model with larger turbulence degree and a non steady evolution of the system. We consider only steady state simulations with $R_{e} = 200$ in the parametric study to save computational time due to the large number of different models analyzed. The authors demonstrated in previous communications \citep{PhysRevE.92.063015,Varela2017} that simulations with $R_{e} = 200$ show the same trends than turbulent simulation with $R_{e} = 1000$. In addition, $R_{e} = 200$ simulations are more didactic to explain the physical mechanisms, avoiding the complexity of turbulent $R_{e} = 1000$ simulations with precession vortex. The robustness of the model was further tested for different values of magnetic diffusivity \citep{Varela2017} ($\lambda$) for a range of simulations with magnetic Reynolds number $R_{m}= V l/\lambda$ between $0.1$ to $10^4$. The effective magnetic Reynolds number of the numerical magnetic diffusion $\eta$ due to the model resolution corresponds to $R_{m} = V L/\eta \approx 6 \cdot 10^{3}$ in the parametric analysis although in the high resolution models the value of the magnetic diffusivity is selected to have $R_{m} =100$ ($P_{m}=R_{m}/R_{e}=0.1$). It should be noted that in the experiment the kinetic Reynolds number can reach $5 \cdot 10^{5}$ and the magnetic Reynolds number is about 50 (for liquid sodium at $120^o$C), so the magnetic Prandtl number is about $P_{m} = 10^{-5}$. A system with such kinetic Reynolds numbers is above the present numerical capabilities by several orders of magnitude without a turbulence model.

\begin{figure}[h]
\centering
\includegraphics[width=0.5\columnwidth]{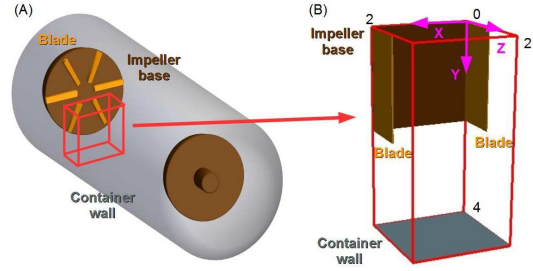} 
\caption{(A) Schematic representation of the VKS experiment geometry using straight blades, (B) simulation domain for a portion of the flow in between two blades: X,Y and Z directions correspond to local azimuthal, radial and vertical directions respectively with X $\in [0, 2]$, Y$ \in [0, 4]$ and Z $\in [0, 2]$.}
\label{1}
\end{figure}

Several of the diagnostics used in the study are quantities averaged in a volume nearby the whirl ($[A] =\int A dxdydz/\int dxdydz$) such as the kinetic energy $[KE]=[\rho v^2 / 2]$, the magnetic energy $[ME]=[B^2/(2\mu_{s})]$ (with $\mu_{s}$ the magnetic permeability of the sodium), the kinetic helicity $[KH] = [\vec{v} \cdot \vec{\omega}]$ (with $\vec{\omega} = \vec{\nabla} \times \vec{v}$ the vorticity), the current helicity $[JH] = [ \vec{B} \cdot \vec{J}]$ (with $ \vec{J} = (\vec{\nabla} \times \vec{B})/\mu_{s}$ the current density) and the total helicity $[He_{T}] = [JH] - [KH]$. In addition, we analyze fluctuating quantities such as the kinetic helicity of the fluctuations $[KH_{f}]$, the current helicity $[JH_{f}]$ and the total helicity $[He_{f}]$ as $[He_f]= [\vec{B'}\cdot\vec{J'}/\rho - \vec{v'}\cdot\vec{\omega'}]$ where the~$'$ denotes the fluctuating part with respect to the time-average ($A' = A -  \langle A \rangle $).

In the following, we consider as reference case the model with azimuthal RMF orientation of intensity $10^{-3}$ T, $\Gamma = 0.8$, $L = 1.0$ m and $D = 2$ m. This model is identified as the TM 73 impeller configuration rotating in the unscooping direction. All the models analyzed are summarized in the appendix ~~\texttt{Model summary}.

\section{Effect of the remnant magnetic field orientation \label{sec:RMF}}

We have run several MHD simulations with the RMF oriented in the local azimuthal $\vec{X}$, radial $\vec{Y}$ and vertical $\vec{Z}$ direction. There is a generation of a magnetic field in the radial (Y) direction and the magnetic field lines (Fig 2, black lines) are collimated by the whirl (Fig 2, green lines) leading to a local enhancement of the magnetic field (Fig 2, pink isocontour). The largest magnetic field is observed in the model with radial RMF orientation (Fig 2B) and the smallest for the azimuthal orientation (Fig 2A), so the radial orientation leads to the most efficient collimation of the magnetic field by the flows.

\begin{figure}[h]
\includegraphics[width=0.5\columnwidth]{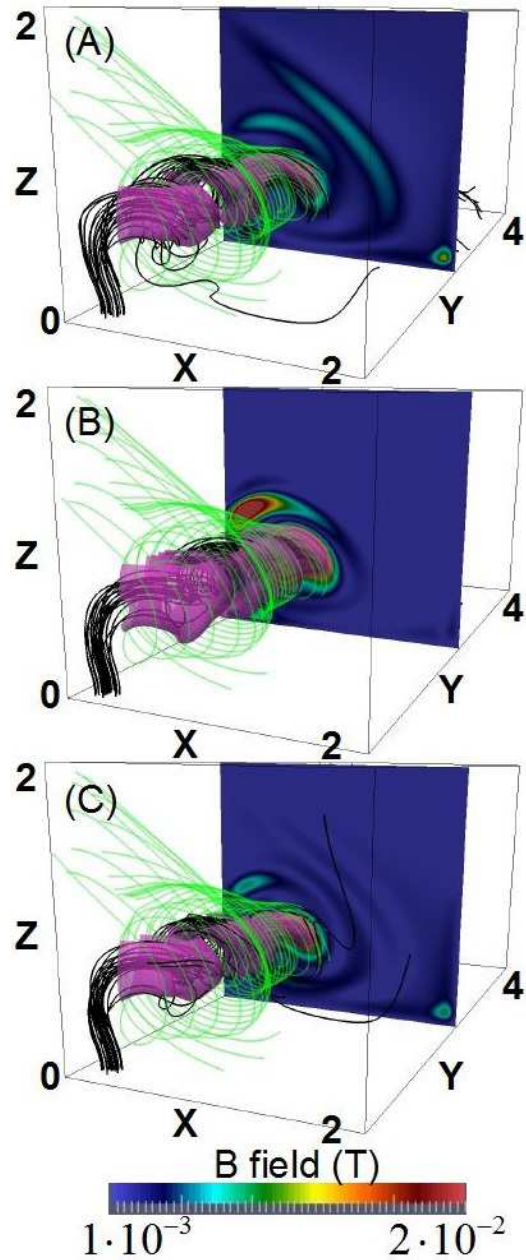} 
\caption{(Color online) Magnetic (black) and velocity (green) field stream lines for a remnant magnetic field oriented in the local azimuthal (A), radial (B) and vertical (C) orientation. Isocontour (pink) of the magnetic field is 0.005 T for the azimuthal orientation and 0.01 T for the other orientations. Magnetic field module is plotted at Y = 2 plane (color bar).}
\label{2}
\end{figure}

Figure 3 shows that regions of alignment of magnetic and velocity fields (blue isocontour indicates angles smaller than $5^o$) are associated with a local maxima of the magnetic energy (color bar, red color) and regions of anti-alignment with a local minima (red isocontour indicates angles larger than $150^o$) inside the whirl. The vertical RMF orientation leads to the angular distribution with the largest regions of anti-alignment (Fig 3C, white arrow), while the azimuthal RMF orientation shows the angular distribution with largest regions of alignment around the whirl (Fig 3A, black arrow). On the other side, the angular distribution with the smallest anti-alignment region nearby the vortex is the case with radial RMF orientation (Fig 3B). The configuration with radial RMF orientation shows the strongest enhancement of the magnetic field, because the model has the most efficient collimation of the magnetic field by the helical flows, due to the presence of a wide region of magnetic and velocity fields alignment nearby the vortex.

\begin{figure}[h]
\includegraphics[width=0.5\columnwidth]{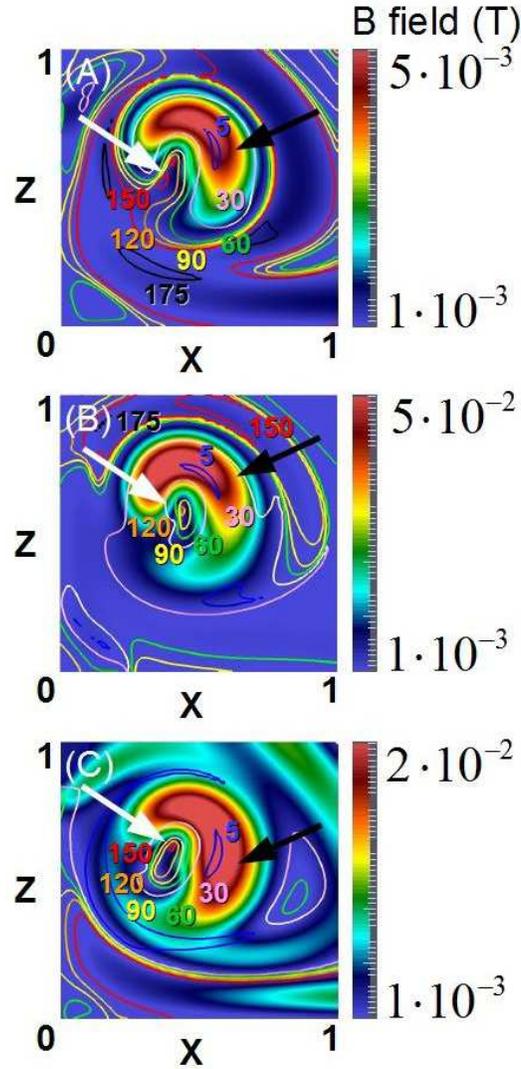} 
\caption{(Color online) Module of the magnetic field (color bar) and angle between velocity and magnetic fields (color isocontours) for a RMF oriented in the azimuthal (A), radial (B) and vertical (C) direction. Plane $Y = 1$ ($t=1.0$ s). The region of parallel alignment between the velocity and the magnetic fields is indicated by a black arrow and the region of anti-alignment by a white arrow.}
\label{3}
\end{figure}

To quantify the enhancement of the magnetic field for the different RMF orientations we use the magnetic energy averaged in a region nearby the whirl (Fig. 4C). The model with radial RMF orientation shows the largest amount of magnetic energy, more than six times larger compared to the vertical RMF orientation and almost 40 times larger compared to the azimuthal RMF case. In models with vertical and radial RMF orientation the magnetic field reaches locally values larger than $10^{-1}$ T ($10^{3}$ gauss), driving an appreciable feedback on the mean flow kinetic helicity (Fig 4A). This effect is not observed in the model with azimuthal orientation because the magnetic field is not strong enough (around $10^{-2}$ T). Consequently, the current helicity is $35$ times larger for the radial RMF orientation compared to the azimuthal RMF orientation and $2$ times larger compared to the vertical RMF orientation (Fig 4B). In all cases the total helicity is dominated by the kinetic term (Fig 4D).

\begin{figure}[h]
\includegraphics[width=0.5\columnwidth]{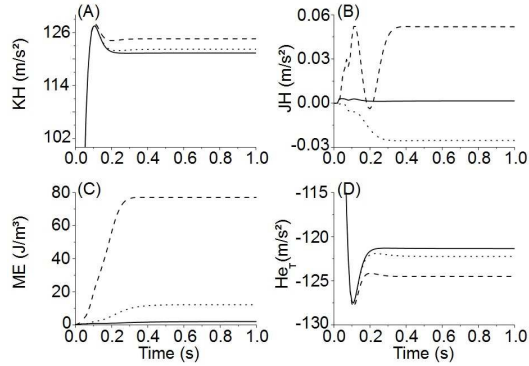} 
\caption{(A) Kinetic helicity, (B) current helicity, (C) magnetic energy and (D) total helicity. The solid line represents the model with local azimuthal orientation of the RMF, the dashed line the radial orientation and the dotted line the vertical orientation. All values are averaged in a volume localized around the whirl.}
\label{4}
\end{figure}

The helicity of the fluctuations is splitted in current $ JH_{f}$ and kinetic $KH_{f}$ terms in figure 5. For all RMF orientations $JH_{f}$ dominates, particularly for the radial RMF orientation that shows the largest total helicity of the fluctuations (Fig 5B), explaining why this model is the most efficient to enhance the magnetic energy of the system: it is the model more sensitive to the ferromagnetic boundary condition. In the following, we only show the values of the fluctuations in the saturated regime for simplicity.

\begin{figure}[h]
\includegraphics[width=0.5\columnwidth]{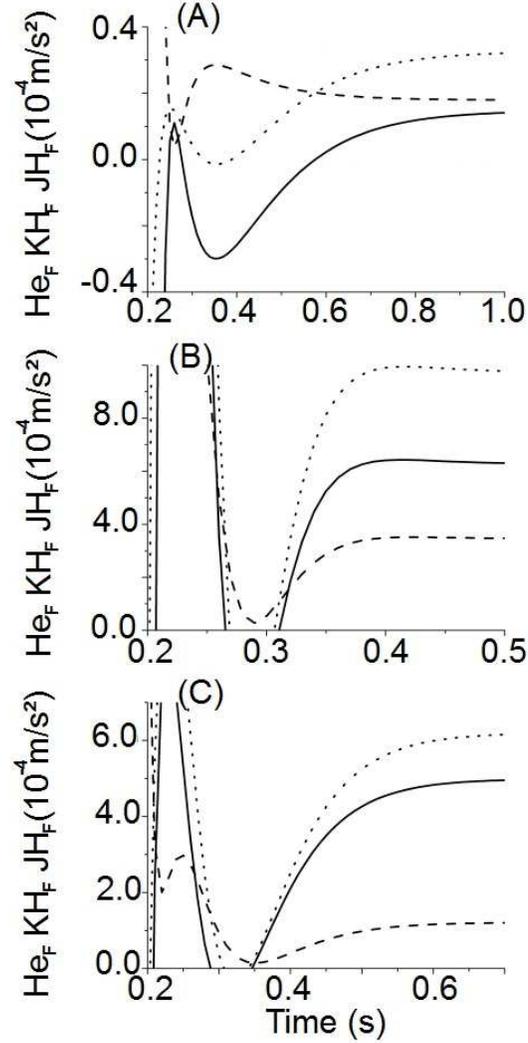} 
\caption{Helicity of fluctuations: total (solid line), kinetic (dashed line) and current (dotted line) fluctuating helicity for (A) azimuthal, (B) radial and (C) vertical RMF orientations. All values are averaged in a volume localized around the whirl.}
\label{5}
\end{figure}

A pure radial RMF orientation is not possible in VKS experiments, it is mainly oriented in the azimuthal orientation. The azimuthal RMF orientation is the less sensitive to the ferromagnetic boundary conditions, so an improvement of the magnetic field collimation requires a RMF orientation with larger components in the vertical and radial (poloidal) directions.

\section{Effect of the remnant magnetic field intensity \label{sec:RMF_I}}

In this section we perform a new set of simulations varying the intensity of the RMF between the values $B0/2$, $2B0$, $5B0$ and $10B0$, analyzing the effect  of the RMF intensity on the magnetic field collimation by the helical flows.

The study shows that switch the RMF intensity doesn't modify the distribution of aligned/anti-aligned regions of magnetic and velocity fields for a configuration with the RMF oriented in the azimuthal direction (Fig 6). This points out that the magnetic field is not large enough to affect the bulk flows, so the collimation of the magnetic field by the flows is almost the same. The magnetic field is enhanced because the available magnetic energy of the system is larger. 

\begin{figure}[h]
\includegraphics[width=0.5\columnwidth]{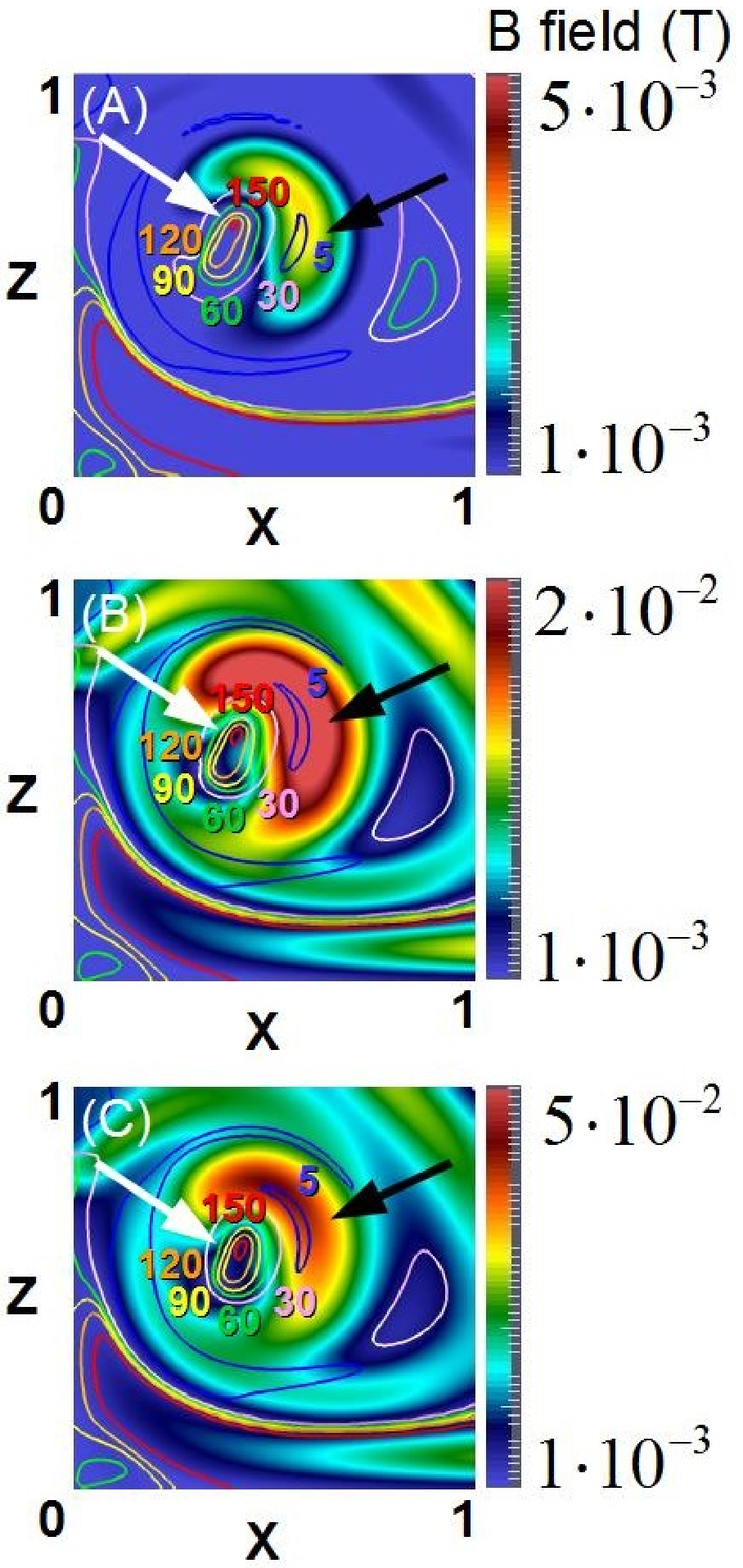} 
\caption{(Color online) Module of the magnetic field (color bar) and angle between velocity and magnetic fields (color isocontours) if the RMF intensity is half (A), 5 times (B) and 10 times (C) the reference RMF intensity. Plane $Y = 1$ ($t=1.0$ s). The region of parallel alignment between the velocity and the magnetic fields is indicated by a black arrow and the region of anti-alignment by a white arrow.}
\label{6}
\end{figure}

$KH$ (Fig 7A), $JH$ (Fig 7B) and magnetic energy (Fig 7C) increase with the RMF intensity. The total helicity (Fig 7D) also increases with the RMF intensity except for models with radial RMF orientation and intensities larger than $5 \cdot 10^{-3}$ T, because the $JH$ grows faster than the $KH$ leading to a drop of the total helicity, indicating that the magnetic field enhancement is large enough to quench the bulk flows. The magnetic energy of the system scales with the RMF intensity as $ME \approx B_{0}^{1.68}$ for a azimuthal RMF orientation, $ME \approx B_{0}^{1.50}$ for the a radial RMF orientation and $ME \approx B_{0}^{1.25}$ for a vertical RMF orientation. Thus the azimuthal RMF orientation is the most efficient to increase the magnetic energy of the system by enhancing the intensity of the RMF.  

\begin{figure}[h]
\includegraphics[width=0.5\columnwidth]{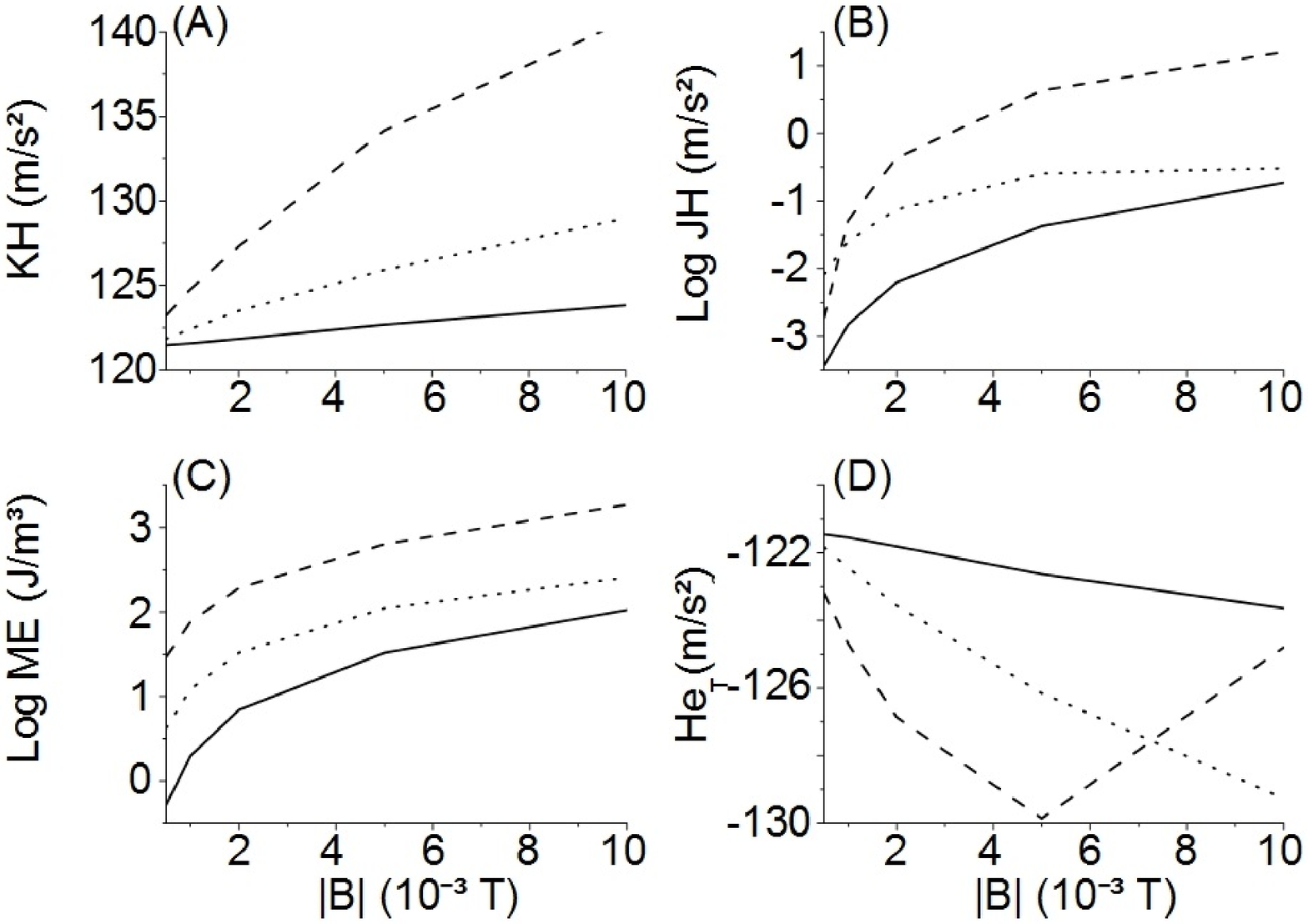} 
\caption{Dependency of the (A) kinetic helicity, (B) current helicity, (C), magnetic energy and (D) total helicity with the RMF intensity. The solid line represents the model with local azimuthal orientation of the RMF, the dashed line the radial and the dotted line the vertical orientation. All values are averaged in a volume localized around the whirl.}
\label{7}
\end{figure}

The helicity of the fluctuations for the models with azimuthal RMF orientation shows a drop of the $He_{f}$ if the RMF intensity increases (Table I), although $He_f$ grows if the RMF intensity decreases, pointing out that the magnetic field collimation efficiency decreases if the RMF intensity increases. The reason is the decrease of the $JH_{f}$ while the $KH_{f}$ remains almost constant (except in the simulation with 10 times the RMF intensity, because the magnetic field is large enough to affect the bulk flow, reducing slightly the $KH$). The total helicity of the fluctuations is 5 times larger for the models with radial RMF if the intensity is 10 times higher, and the $KH_{f}$ enhancement is larger compared to the $JH_{f}$. In the models with vertical RMF and 10 times the RMF intensity, the total helicity of the fluctuation decreases due to the drop of the $KH_{f}$ and $JH_{f}$.

\begin{table}[h]
\centering
\begin{tabular}{ c | c  c  c }
Model & $He_f$ & $KH_{f}$  & $JH_{f}$ \\
\hline
$BX_{B0/2}$ & $0.22$ & $0.22$ & $0.44$ \\
$BX_{5B0}$ & $0.11$ & $0.13$ & $0.24$ \\
$BX_{10B0}$ & $-0.05$ & $0.08$ & $0.03$ \\
$BY_{10B0}$ & $-25$ & $9$ & $-16$ \\
$BZ_{10B0}$ & $0.09$ & $0.19$ & $0.28$ \\
\hline
\end{tabular}
\caption{Helicity of fluctuations (first column), kinetic (second column) and current (third column) fluctuating helicity for models with azimuthal RMF orientations and intensities of half, 5 times and 10 times the reference case, as well as models with radial and vertical RMF orientations and an intensity 10 times the reference case. All values are averaged in a volume localized around the whirl. Table units: $10^{-4}$ m/s$^{2}$.}
\label{1}
\end{table}

In summary, increasing the magnetic energy of the system by enhancing the RMF intensity is less efficient as the RMF intensity increases.

\section{Effect of the impinging velocity \label{sec:impinging_vel}}

In this section we study the effect of the Ekman velocity in the magnetic field collimation, performing simulations with different impinging velocities.

The analysis points out that reduce the impinging velocity due to Ekman pumping, model with $\Gamma = 0.7$, leads to a slight increase of the magnetic field collimation (Fig 8A), correlated with a small reduction of the velocity and magnetic fields anti-alignment region nearby the whirl vortex. Increasing the impinging velocity, model with $\Gamma = 0.9$, leads to a large reduction of the magnetic field collimation (Fig 8B) and a large region of anti-alignment nearby the whirl vortex.  

\begin{figure}[h]
\includegraphics[width=0.5\columnwidth]{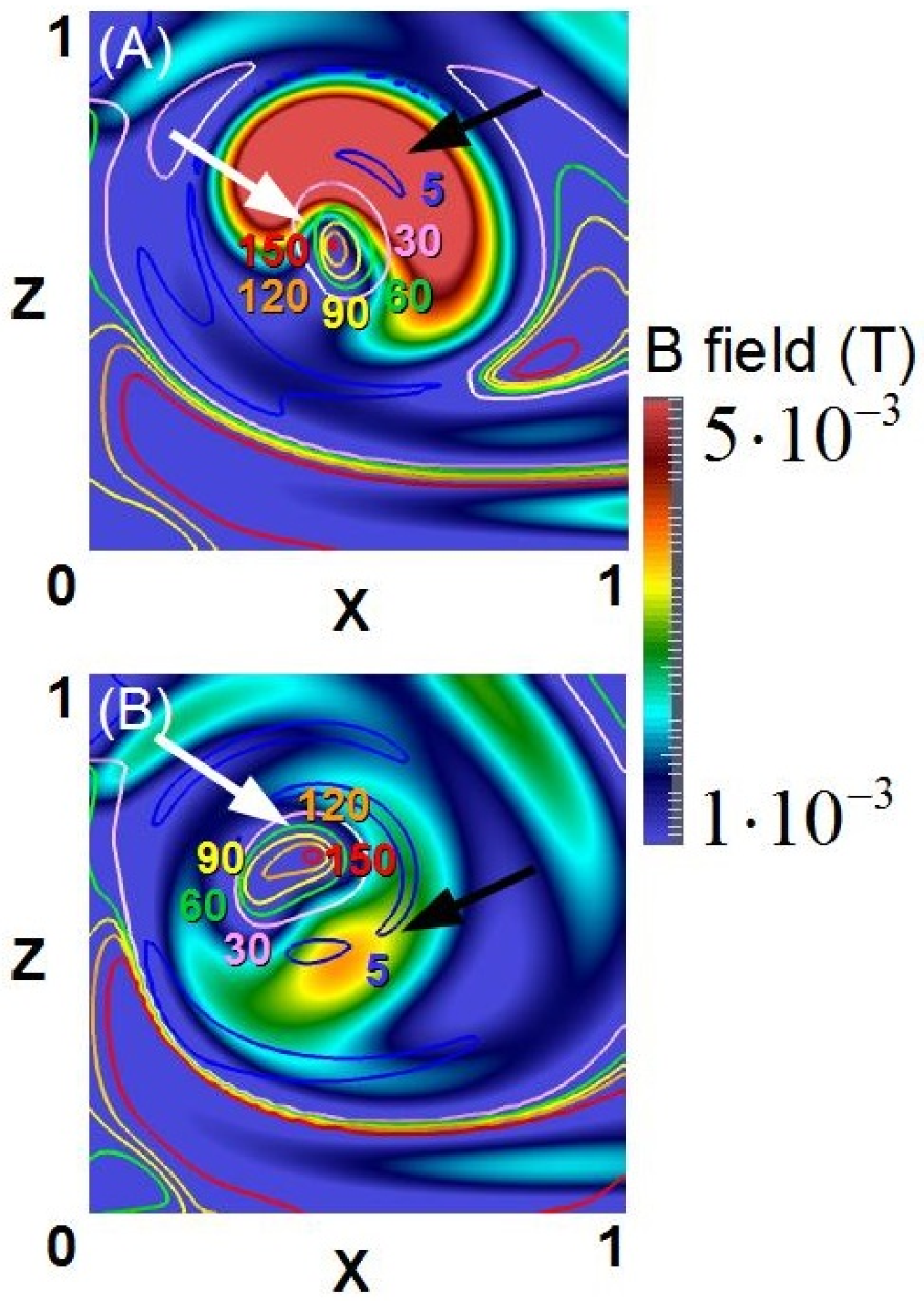} 
\caption{(Color online) Module of the magnetic field (color bar) and angle between the velocity and the magnetic fields (color isocontours) for the models with $\Gamma = 0.7$ (A) and $0.9$ (B) and azimuthal RMF. Plane $Y = 1$ ($t=0.7$ s). The region of parallel alignment between the velocity and the magnetic fields is indicated by a black arrow and the region of anti-alignment by a white arrow.}
\label{8}
\end{figure}

For all RMF orientations the $KH$ (Fig 9A) increases with $\Gamma$ and the $JH$ (Fig 9B) decreases. The total helicity (Fig 9D) is dominated by the $KH$ term. The magnetic energy (Fig 9C) decreases for all $\Gamma = 0.9$ models although it is almost the same for $\Gamma= 0.7$ and $0.8$ models, because the $KH$ decrease in $\Gamma= 0.7$ models is compensated by an enhancement of the $JH$.

\begin{figure}[h]
\includegraphics[width=0.5\columnwidth]{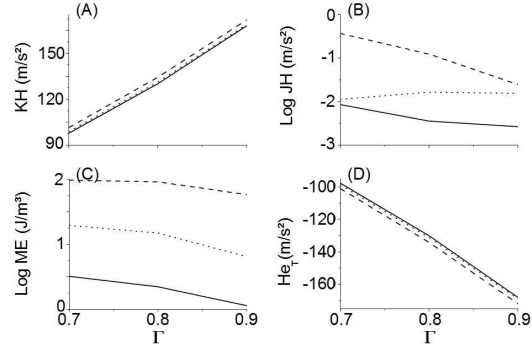} 
\caption{Dependency of (A) kinetic helicity, (B) current helicity, (C), magnetic energy and (D) total helicity with $\Gamma$. The solid line represents the model with local azimuthal orientation of the RMF, the dashed line the radial and the dotted line the vertical orientation. All values are averaged in a volume localized around the whirl.}
\label{9}
\end{figure}

The $He_f$ decreases if $\Gamma$ increases and vice versa (Table II). For azimuthal and vertical RMF orientations, reduce the impinging velocity leads to a decrease of the $He_f$, although for a radial RMF orientation $He_f$ is enhanced. Increase $\Gamma$ leads to a reduction of the ratio between $KH_{f}$ and $JH_{f}$. The $KH_{f}$ dominates over the $JH_{f}$ in the models with azimuthal and vertical RMF orientations.

\begin{table}[h]
\centering
\begin{tabular}{ c | c  c  c }
Model & $He_f$ & $KH_{f}$ & $JH_{f}$ \\
\hline
$BX_{\Gamma=0.7}$ & $-0.37$ & $0.55$ & $0.18$ \\
$BX_{\Gamma=0.9}$ & $-0.24$ & $0.06$ & $-0.18$ \\
$BY_{\Gamma=0.7}$ & $15.5$ & $10.1$ & $25.4$ \\
$BY_{\Gamma=0.9}$ & $1.12$ & $0.25$ & $1.37$ \\
$BZ_{\Gamma=0.7}$ & $6.42$ & $-0.11$ & $6.31$ \\
$BZ_{\Gamma=0.9}$ & $-0.01$ & $0.55$ & $0.54$ \\
\hline
\end{tabular}
\caption{Helicity of the fluctuations (first column), kinetic (second column) and current (third column) fluctuating helicity for models with azimuthal, radial and vertical RMF orientations with $\Gamma = 0.7$ and $0.9$. All values are averaged in a volume localized around the whirl. Table units: $10^{-4}$ m/s$^{2}$.}
\label{2}
\end{table}

The enhancement of the impinging velocity for a ratio larger than $\Gamma = 0.8$ leads to a decrease of the magnetic field collimation, however decrease $\Gamma$ only improves slightly the collimation of models with azimuthal and vertical RMF orientations, pointing out that the optimal configuration must be in between $\Gamma = 0.8$ and $0.7$. 

The next step of the study is to analyze the magnetic field collimation by impellers with different blade and base dimensions.

\section{Effect of the blade height \label{sec:height}}

In this section we study the effect of the blade height in the magnetic field collimation.

The analysis shows that decrease the blade height leads to a demise of the anti-alignment region nearby the whirl vortex (Fig 10A). The whirl is more focused and localized so the magnetic field collimation is enhanced. Increasing the blade height yields the opposite scenario, the region of anti-alignment nearby the whirl vortex increases (Fig 10B), the whirl is wider, the magnetic field enhancement is weaker and the local maximum is located around the whirl.

\begin{figure}[h]
\includegraphics[width=0.5\columnwidth]{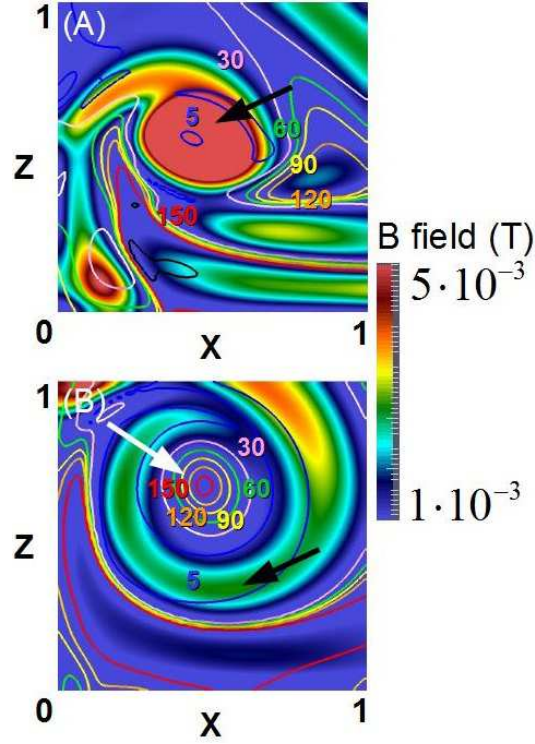}
\caption{(Color online) Module of the magnetic field (color bar) and angle between velocity and magnetic fields (color isocontours) for models with a blade height of $L = 0.75$ m (A) and $L = 1.25$ m (B) for an azimuthal RMF. Plane $Y = 1$ ($t=0.7$ s). The region of parallel alignment between the velocity and the magnetic fields is indicated by a black arrow and the region of anti-alignment by a white arrow.}
\label{10}
\end{figure}

Reducing or increasing the blade height leads to a drop of the magnetic energy for all RMF orientations except for the azimuthal orientation, showing a slight increase of the magnetic energy if the blade height increases (Fig 11C). For all RMF orientations the configuration with $L = 0.75$ m shows the largest total helicity and the configuration with $L = 1.25$ m the smallest (Fig 11D). The models with azimuthal RMF orientation have the largest increment of the $JH$ if the blade height decreases from $L = 1.0$ to $0.75$ m, almost two order of magnitude larger, although for the radial and vertical RMF orientations is only the double (Fig 11B). The variation of the $KH$ is similar for all RMF orientations (Fig 11A).

\begin{figure}[h]
\includegraphics[width=0.5\columnwidth]{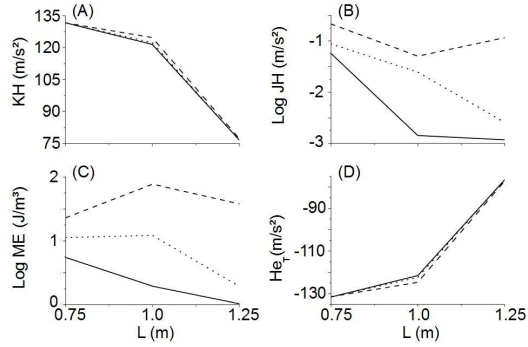} 
\caption{Dependency of the (A) kinetic helicity, (B) current helicity, (C), magnetic energy and (D) total helicity with the blade height. The solid line represents the model with local azimuthal orientation of the RMF, the dashed line the radial and the dotted line the vertical orientation. All values are averaged in a volume localized around the whirl.}
\label{11}
\end{figure}

The helicity of the fluctuations is slightly larger on the model with azimuthal RMF orientation if the blade height decreases (Table III) although the $JH_{f}$ is weaker, dominated by the $KH_{f}$ and leading to a configuration less sensitive to the ferromagnetic boundary conditions. For all the other RMF orientations and blade heights, the helicity of the fluctuation is smaller and dominated by the $KH_{f}$, except for the model with radial RMF orientation and a blade height of $L = 0.75$, which $JH_{f}$ is still dominant.   

\begin{table}[h]
\centering
\begin{tabular}{ c | c  c  c }
Model & $He_f$ & $KH_{f}$ & $JH_{f}$ \\
\hline
$BX_{L=0.75}$ & $-0.31$ & $0.33$ & $0.02$ \\
$BX_{L=1.25}$ & $-0.07$ & $0.07$ & $0.00$ \\
$BY_{L=0.75}$ & $0.81$ & $0.29$ & $1.10$ \\
$BY_{L=1.25}$ & $0.01$ & $0.87$ & $0.88$ \\
$BZ_{L=0.75}$ & $-0.10$ & $0.27$ & $0.17$ \\
$BZ_{L=1.25}$ & $-0.11$ & $0.14$ & 0.03$$ \\
\hline
\end{tabular}
\caption{Helicity of the fluctuations (first column), kinetic (second column) and current (third column) fluctuating helicity for models with azimuthal, radial and vertical RMF orientations with $L=0.75$ and $1.25$ m. All values are averaged in a volume localized around the whirl. Table units: $10^{-4}$ m/s$^{2}$.}
\label{3}
\end{table}

The most efficient ratio between blade height and impeller base length (fixed) to enhance the magnetic energy of the system is $0.5$ for all RMF orientations except the azimuthal case. For the azimuthal orientation the optimal ratio is in between $0.5$ and $0.375$. 

\section{Effect of the base length \label{sec:base}}

We conclude the optimization analysis studying the effect of the base length on the magnetic field collimation.

Compared to the reference case the study indicates that the model with an impeller base length of $D = 1$ m shows a weaker enhancement of the magnetic field and the whirl is located closer to the left impeller blade (Fig 12A). No anti-alignment region nearby the whirl vortex is observed. The model with an impeller base length of $1.5$ m has a similar magnetic field enhancement and angle distribution regarding to the reference case (Fig 12B). The model with $D = 2.5$ m shows a smaller magnetic field and a wider whirl with a larger region of anti-alignment nearby the vortex compared to the reference case (Fig 12C).

\begin{figure}[h]
\includegraphics[width=0.5\columnwidth]{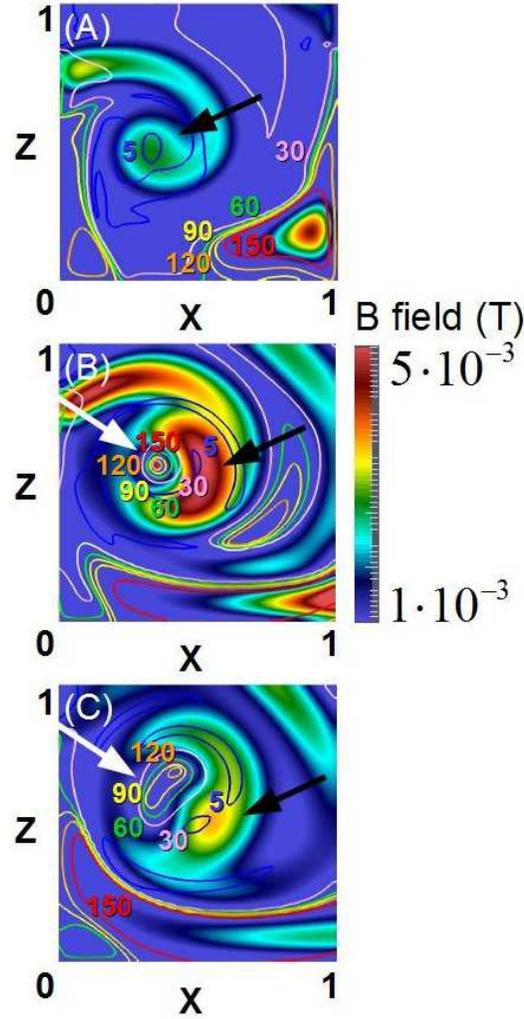} 
\caption{(Color online) Module of the magnetic field (color bar) and angle between velocity and magnetic fields (color isocontours) for models with impeller base length of $D=1.0$ m (A), $1.5$ m (B) and $2.5$ m (C) and a azimuthal RMF. Plane $Y = 1$ ($t=0.7$ s). The region of parallel alignment between the velocity and the magnetic fields is indicated by a black arrow and the region of anti-alignment by a white arrow.}
\label{12}
\end{figure}

For all RMF orientations except the azimuthal RMF orientation, the configurations with the largest magnetic energy are the models with an impeller base length of $D = 2.0$ m (Fig 13C). For the azimuthal RMF orientation the magnetic energy is slightly larger if $D = 1.5$ m. Reduce the impeller base length leads to a decrease of the $KH$ (Fig 13A), however the configurations with the largest $JH$ are the models with an impeller length of $D = 1.5$ m (Fig 13B), except for the cases with vertical RMF orientation that show the largest $JH$ if $D = 2.0$ m. The total helicity increases with the impeller length (Fig 13D), although it is similar for $D = 2.0$ and $D = 2.5$ m configurations because the $KH$ is smaller and the $JH$ decreases.

\begin{figure}[h]
\includegraphics[width=0.5\columnwidth]{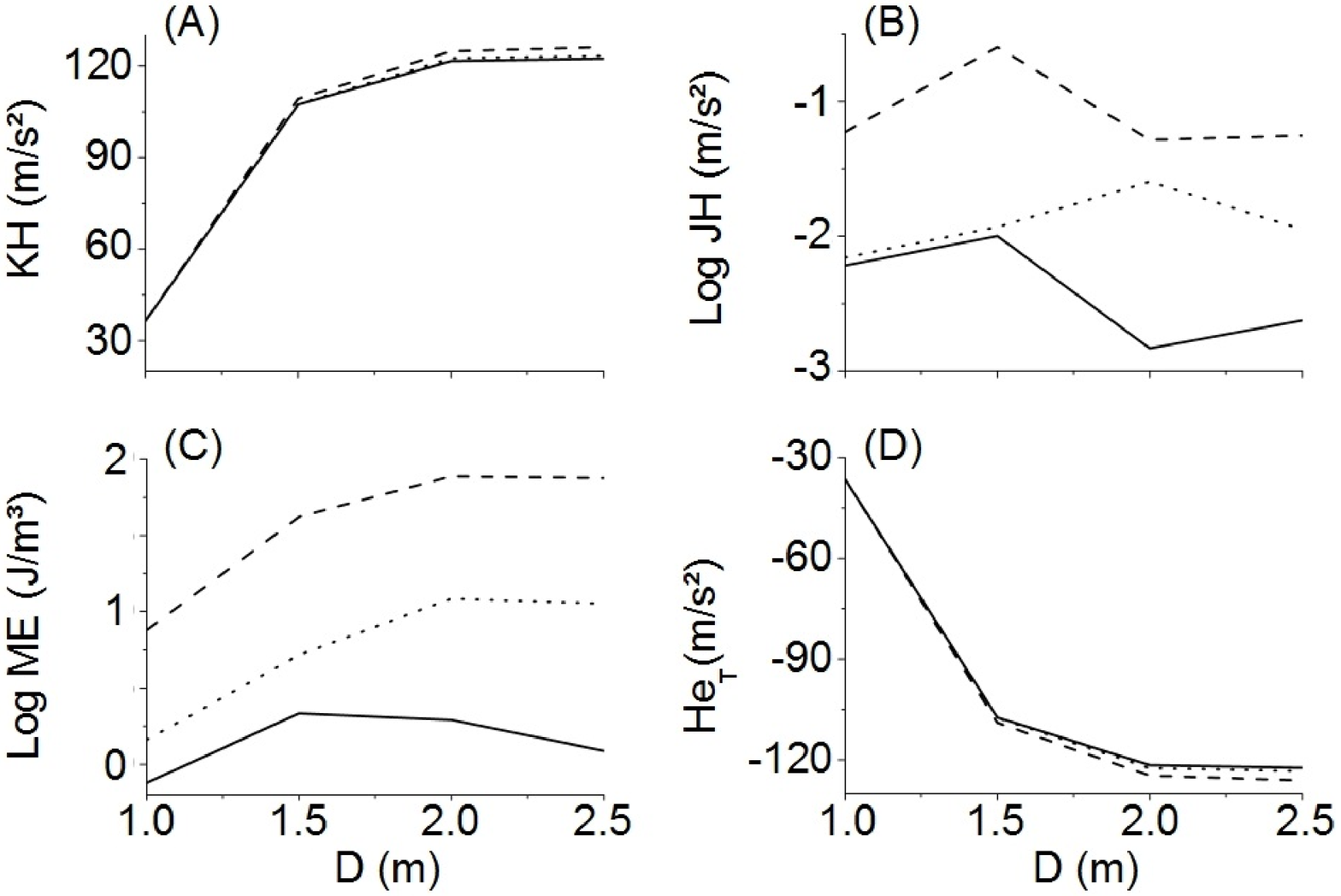} 
\caption{Dependency of the (A) kinetic helicity, (B) current helicity, (C), magnetic energy and (D) total helicity with the impeller base length. The solid line represents the model with local azimuthal RMF orientation, the dashed line the radial and the dotted line the vertical orientation. All values are averaged in a volume localized around the whirl.}
\label{13}
\end{figure}

The $He_{f}$ decreases with the impeller base length, similar for $D = 2.5$ and $D = 2.0$ m models (Table IV). The $KH_{f}$ is larger than the $JH_{f}$ in the model $D = 2.5$ m, leading to a weaker influence of the ferromagnetic boundary conditions and a lower enhancement of the magnetic fields. The same trend is observed for other orientations, although the $JH_{f}$ term is still dominant if the impeller base length is $D = 1.5$ m for radial and vertical RMF orientations.

\begin{table}[h]
\centering
\begin{tabular}{ c | c  c  c }
Model & $He_f$ & $KH_{f}$ & $JH_{f}$ \\
\hline
$BX_{D=1.5}$ & $-0.08$ & $0.08$ & $0.00$ \\
$BX_{D=2.5}$ & $-0.13$ & $0.31$ & $0.18$ \\
$BY_{D=1.5}$ & $1.77$ & $0.14$ & $1.91$ \\
$BY_{D=2.5}$ & $0.41$ & $0.2.08$ & $2.49$ \\
$BZ_{D=1.5}$ & $0.05$ & $0.04$ & $0.09$ \\
$BZ_{D=2.5}$ & $-1.32$ & $2.94$ & $1.62$ \\
\hline
\end{tabular}
\caption{Helicity of the fluctuations (first column), kinetic (second column) and current (third column) fluctuating helicity for models with impeller base length of $D=1.0$ m, $1.5$ m and $2.5$ m for an azimuthal RMF. All values are averaged in a volume localized around the whirl. Table units: $10^{-4}$ m/s$^{2}$.}
\label{4}
\end{table}

The analysis indicates that the optimum ratio between blade height (fixed) and impeller base length to enhance the magnetic field collimation is $0.5$ for all RMF orientations, except the azimuthal orientation whose optimal ratio is in between $0.5$ and $0.375$.

\section{Dynamo loop \label{sec:dynamo}}

We further quantify the efficiency of the models magnetic field collimation analyzing the 9 components of the helicity tensor $h_{ij}$ (data not shown), defined as:

\begin{equation}
 h_{ij} = \epsilon_{ikn} \langle  u_{k}^{'} \partial_{j} u_{n}^{'} \rangle 
\label{1}
\end{equation}
 We use the helicity tensor to estimate the classical mean field dynamo mechanisms occurring in the vicinity of the impeller. We quantify the hypothetical $\alpha^{2}$ dynamo loop based on the toroidal magnetic field regeneration ($B_{x}$) from the poloidal magnetic field ($B_{y}$ and $B_{z}$) through helicoidal motion. The main dynamo loops are: 

\begin{eqnarray} 
B_y \xrightarrow[\text{}]{\text{hyy,hzy}} B_x \xrightarrow[\text{}]{\text{hxx,hzx}} B_y \\
B_z \xrightarrow[\text{}]{\text{hzz,hyz}} B_x \xrightarrow[\text{}]{\text{hyx,hxx}} B_z
\end{eqnarray} 
The toroidal imposed velocity field experiences a vertical shear in the vicinity of the impeller that can also regenerate the $B_x$ component from $B_z$, resulting in an $\Omega-\alpha$ dynamo loop. The hypothetical $\Omega$-$\alpha$ dynamo loop is defined as: 

\begin{equation}
B_z \stackrel{\Omega'}{\longrightarrow} B_x \xrightarrow[\text{}]{\text{hyx,hxx}} B_z
\end{equation}
with $\Omega^{'} = \partial <u_{x}> / \partial z = (<u_{x}>_{top} - <u_{x}>_{bottom})/L_{blade})$, $L_{blade}$ the blade height, $<u_{x}>_{top}$ the azimuthal averaged velocity at the top of the impeller and $<u_{x}>_{bottom} = 0$ the velocity at the bottom of the impeller, traduced in the products: $(\Omega{'} h_{yx})$ and $(\Omega{'} h_{xx})$.

To assess the study of the configurations that leads to the largest dynamo loop enhancement we define the gain factor $G$, the ratio between the dynamo loop components of the model and the dynamo loop components of the reference case (TM 73). The gain factor is defined for the $\alpha^{2}$ dynamo loop as:

\begin{equation}
G_{ij,km} = \frac{|\langle h_{ij} h_{km} \rangle|_{model}} {|\langle h_{ij} h_{km} \rangle|_{TM 73}}
\label{2}
\end{equation}
and for the  $\Omega$-$\alpha$ dynamo loop as: 

\begin{equation}
G_{im} = \frac{|\langle \Omega^{'} h_{im} \rangle|_{model}}  {|\langle \Omega^{'} h_{km} \rangle|_{TM 73}}
\label{3}
\end{equation}
To identify the dominant dynamo loop we compute the autocorrelation time $C_{\tau}$ and the autocorrelation distance $C_{d}$ of the mean velocity, because from the dimensional analysis:

\begin{equation}
[h_{ij} h_{km} \delta_{jk}] = \left[ \frac{C_{d}}{C_{\tau}} \Omega^{'} h_{im} \right]
\label{4}
\end{equation}
A detailed definition of the autocorrelation functions of time and distance is included in the appendix~~\texttt{Autocorrelation}. The autocorrelation factor is $C_{d} / C_{\tau} \approx 0.25$ m/s for the reference case. In tables \ref{5} and \ref{6} we show the gain factor of the hypothetical $\alpha^{2}$ dynamo loop. The table \ref{7} (columns 1 and 2) shows the gain factor of the hypothetical $\Omega$-$\alpha$ dynamo loop. The radial ($BY_{B0}$ model) and vertical ($BZ_{B0}$ model) RMF orientations lead to an enhancement of the hypothetical $\alpha^{2}$ and $\Omega$-$\alpha$ dynamo loops, particularly for the radial orientation. If the RMF intensity in the toroidal direction is reduced by half of the reference case intensity ($BX_{B0/2}$ model) there is a drop of almost two orders of magnitude in the $\alpha^{2}$ dynamo loop components, although the components of the $\Omega$-$\alpha$ dynamo loop are nearly the same. Increase the RMF intensity yields an enhancement of the dynamo loop components only for the $BX_{5B0}$ model, but not for the $BX_{10B0}$ model, pointing out that the efficiency of $\alpha^{2}$ and $\Omega$-$\alpha$ dynamo loops drop if the RMF intensity increases. The hypothetical $\alpha^{2}$ dynamo loop is weaker if the RMF intensity increases for a RMF oriented in the vertical direction ($BZ_{10B0}$ model), although the hypothetical $\Omega$-$\alpha$ dynamo loop remains almost the same. If the intensity of a RMF oriented in the radial direction increases, both dynamo loops are enhanced more than one order of magnitude ($BY_{10B0}$ model). $\Gamma_{0.7}$ model shows a slightly enhancement of the $\alpha^{2}$ and $\Omega$-$\alpha$ dynamo loops, although $\Gamma_{0.9}$ model leads to a weaker $\alpha^{2}$ (almost two orders of magnitude smaller compared to the reference case) and $\Omega$-$\alpha$ dynamo loops (half of the reference case). Reducing the height of the blades ($L_{0.75}$ model) leads to a slightly weaker $\alpha^{2}$ dynamo loop and a small enhancement of the $\Omega$-$\alpha$ dynamo loop, although both dynamos loops are strongly weakened if the blade height increases ($L_{1.25}$ model). Reducing or increasing the impeller base length leads to a weaker $\alpha^{2}$ dynamo loop and a small enhancement of the $\Omega$-$\alpha$ dynamo loop.   

\begin{table}[h]
\centering

\begin{tabular}{c}
$BY_{B0}$ \\
\end{tabular}

\begin{tabular}{c | c | c | c }
\hline
$ G_{yy,zx} $ = 0.80 & $G_{yy,xx} $ = 21.7 & $ G_{zy,xx} $ = 50.3 & $ G_{zy,zx} $ = 1.86 \\
$ G_{zz,yx} $ = 46.3 & $G_{zz,xx} $ = 30.1 & $ G_{yz,yx} $ = 60.2 & $ G_{yz,xx} $ = 39.1 \\
\hline
\end{tabular}

\begin{tabular}{c}
$BZ_{B0}$ \\
\end{tabular}

\begin{tabular}{c | c | c | c }
\hline
$ G_{yy,zx} $ = 16.1 & $G_{yy,xx} $ = 58.1 & $ G_{zy,xx} $ = 74.0 & $ G_{zy,zx} $ = 20.5 \\
$ G_{zz,yx} $ = 99.5 & $G_{zz,xx} $ = 77.5 & $ G_{yz,yx} $ = 28.0 & $ G_{yz,xx} $ = 21.8 \\
\hline
\end{tabular}

\begin{tabular}{c}
$BX_{B0/2}$ \\
\end{tabular}

\begin{tabular}{c | c | c | c }
\hline
$ G_{yy,zx} $ = 0.01 & $G_{yy,xx} $ = 0.04 & $ G_{zy,xx} $ = 0.02 & $ G_{zy,zx} $ = 0.01 \\
$ G_{zz,yx} $ = 0.04 & $G_{zz,xx} $ = 0.02 & $ G_{yz,yx} $ = 0.42 & $ G_{yz,xx} $ = 0.25 \\
\hline
\end{tabular}

\begin{tabular}{c}
$BX_{5B0}$ \\
\end{tabular}

\begin{tabular}{c | c | c | c }
\hline
$ G_{yy,zx} $ = 1.47 & $G_{yy,xx} $ = 5.01 & $ G_{zy,xx} $ = 6.36 & $ G_{zy,zx} $ = 1.87 \\
$ G_{zz,yx} $ = 9.25 & $G_{zz,xx} $ = 6.30 & $ G_{yz,yx} $ = 3.82 & $ G_{yz,xx} $ = 2.60 \\
\hline
\end{tabular}

\begin{tabular}{c}
$BX_{10B0}$ \\
\end{tabular}

\begin{tabular}{c | c | c | c }
\hline
$ G_{yy,zx} $ = 0.27 & $G_{yy,xx} $ = 1.99 & $ G_{zy,xx} $ = 2.97 & $ G_{zy,zx} $ = 0.41 \\
$ G_{zz,yx} $ = 3.60 & $G_{zz,xx} $ = 2.37 & $ G_{yz,yx} $ = 3.25 & $ G_{yz,xx} $ = 2.14 \\
\hline
\end{tabular}

\begin{tabular}{c}
$BY_{10B0}$ \\
\end{tabular}

\begin{tabular}{c | c | c | c }
\hline
$ G_{yy,zx} $ = 23.5 & $G_{yy,xx} $ = 149 & $ G_{zy,xx} $ = 136 & $ G_{zy,zx} $ = 21.4 \\
$ G_{zz,yx} $ = 60.2 & $G_{zz,xx} $ = 40.1 & $ G_{yz,yx} $ = 44.5 & $ G_{yz,xx} $ = 29.6 \\
\hline
\end{tabular}

\begin{tabular}{c}
$BZ_{10B0}$ \\
\end{tabular}

\begin{tabular}{c | c | c | c }
\hline
$ G_{yy,zx} $ = 0.05 & $G_{yy,xx} $ = 0.27 & $ G_{zy,xx} $ = 0.40 & $ G_{zy,zx} $ = 0.07 \\
$ G_{zz,yx} $ = 0.14 & $G_{zz,xx} $ = 0.22 & $ G_{yz,yx} $ = 0.43 & $ G_{yz,xx} $ = 0.68 \\
\hline
\end{tabular}

\caption{Gain factor of the hypothetical $\alpha^{2}$ dynamo loop components.}
\label{5}
\end{table}

\begin{table}[h]
\centering

\begin{tabular}{c}
$\Gamma_{0.7}$ \\
\end{tabular}

\begin{tabular}{c | c | c | c }
\hline
$ G_{yy,zx} $ = 4.94 & $G_{yy,xx} $ = 16.1 & $ G_{zy,xx} $ = 1.10 & $ G_{zy,zx} $ = 0.34 \\
$ G_{zz,yx} $ = 0.53 & $G_{zz,xx} $ = 2.43 & $ G_{yz,yx} $ = 1.14 & $ G_{yz,xx} $ = 5.25 \\
\hline
\end{tabular}

\begin{tabular}{c}
$\Gamma_{0.9}$ \\
\end{tabular}

\begin{tabular}{c | c | c | c }
\hline
$ G_{yy,zx} $ = 0.03 & $G_{yy,xx} $ = 0.13 & $ G_{zy,xx} $ = 0.10 & $ G_{zy,zx} $ = 0.03 \\
$ G_{zz,yx} $ = 0.13 & $G_{zz,xx} $ = 0.16 & $ G_{yz,yx} $ = 0.03 & $ G_{yz,xx} $ = 0.04 \\
\hline
\end{tabular}

\begin{tabular}{c}
$L_{0.75}$ \\
\end{tabular}

\begin{tabular}{c | c | c | c }
\hline
$ G_{yy,zx} $ = 0.15 & $G_{yy,xx} $ = 0.52 & $ G_{zy,xx} $ = 0.59 & $ G_{zy,zx} $ = 0.17 \\
$ G_{zz,yx} $ = 0.48 & $G_{zz,xx} $ = 0.12 & $ G_{yz,yx} $ = 2.44 & $ G_{yz,xx} $ = 0.61 \\
\hline
\end{tabular}

\begin{tabular}{c}
$L_{1.25}$ \\
\end{tabular}

\begin{tabular}{c | c | c | c }
\hline
$ G_{yy,zx} $ = 0.00 & $G_{yy,xx} $ = 0.00 & $ G_{zy,xx} $ = 0.35 & $ G_{zy,zx} $ = 0.01 \\
$ G_{zz,yx} $ = 0.01 & $G_{zz,xx} $ = 0.16 & $ G_{yz,yx} $ = 0.01 & $ G_{yz,xx} $ = 0.18 \\
\hline
\end{tabular}

\begin{tabular}{c}
$D_{1.5}$ \\
\end{tabular}

\begin{tabular}{c | c | c | c }
\hline
$ G_{yy,zx} $ = 0.00 & $G_{yy,xx} $ = 0.30 & $ G_{zy,xx} $ = 1.15 & $ G_{zy,zx} $ = 0.02 \\
$ G_{zz,yx} $ = 0.87 & $G_{zz,xx} $ = 0.82 & $ G_{yz,yx} $ = 0.09 & $ G_{yz,xx} $ = 0.09 \\
\hline
\end{tabular}

\begin{tabular}{c}
$D_{2.5}$ \\
\end{tabular}

\begin{tabular}{c | c | c | c }
\hline
$ G_{yy,zx} $ = 0.06 & $G_{yy,xx} $ = 0.51 & $ G_{zy,xx} $ = 2.38 & $ G_{zy,zx} $ = 0.28 \\
$ G_{zz,yx} $ = 1.63 & $G_{zz,xx} $ = 3.63 & $ G_{yz,yx} $ = 1.04 & $ G_{yz,xx} $ = 2.31 \\
\hline
\end{tabular}

\caption{Gain factor of the hypothetical $\alpha^{2}$ dynamo loop components.}
\label{6}
\end{table}

\begin{table}[h]
\centering

\begin{tabular}{c | c c c }
Model & G($\Omega^{'} h_{yx} $) & G($\Omega^{'} h_{xx} $) & P \\
\hline
Reference & 1 & 1 & 174 \\
$BY_{B0}$ & 13.4 & 8.69 & 33.2 \\
$BZ_{B0}$ & 16.4 & 12.80 & 20.8 \\
$BX_{B0/2}$ & 0.90 & 0.53 & 515 \\
$BX_{5B0}$ & 5.09 & 3.46 & 69.1 \\
$BX_{10B0}$ & 3.50 & 2.30 & 122 \\
$BY_{10B0}$ & 17.87 & 11.9 & 39.2 \\
$BZ_{10B0}$ & 0.68 & 1.09 & 237 \\
$\Gamma_{0.7}$ & 1.04 & 4.79 & 128 \\
$\Gamma_{0.9}$ & 0.45 & 0.53 & 477 \\
$L_{0.75}$ & 2.80 & 0.70 & 152 \\
$L_{1.25}$ & 0.02 & 0.56 & 616 \\
$D_{1.5}$ & 1.48 & 1.39 & 273 \\
$D_{2.5}$ & 1.48 & 3.28 & 113 \\
\hline
\end{tabular}

\caption{Gain factor of the hypothetical $\Omega$-$\alpha$ dynamo loop components (columns 1 and 2) and the ratio of the largest components of the $\Omega$-$\alpha$ and $\alpha^{2}$ ($B_z\to B_x \to B_z$) hypothetical dynamo loops (column 3).}
\label{7}
\end{table}

We calculate the ratio between the largest component of the $\Omega$-$\alpha$ and $\alpha^{2}$ dynamo loops defined as (table \ref{3}, column 3):

\begin{equation}
P = \frac{|\langle \Omega^{'} h_{im} \rangle|_{max}} {|\langle h_{ij} h_{km} \delta_{jk} \rangle|_{max}}
\end{equation}

The model $BZ_{B0}$ has the smallest $P$ ratio, followed by $BY_{B0}$ and $BY_{10B0}$ cases. Model $L_{1.25}$ shows the largest $P$ ratio followed by cases $BX_{B0/2}$ and $\Gamma_{0.9}$. In summary, $P$ ratio increases ($\Omega$-$\alpha$ dynamo loop is reinforced) if: the RMF intensity is enhanced, $\Gamma$ value increases to $0.9$, the blade height is higher or the impeller base length decreases (smaller distance between blades). $P$ ratio drops if: the RMF is oriented in vertical or radial directions, $\Gamma$ value is reduced to $0.7$, the blade height decreases or the impeller base length increases (larger distance between blades). It should be noted that the dominant mechanism in all the simulations is the $\Omega-\alpha$ dynamo loop,  one to two orders of magnitude larger than the $\alpha^2$ dynamo loop.

\section{Optimization study \label{sec:optimization}}

In this section we perform simulations with higher resolution and larger turbulence degree to confirm the optimization trends obtained by the dynamo loop study if the system evolution is non stationary. We study two configurations: optimized model (set up with parameters that leads to an improved magnetic field collimation compared to the reference case) and non optimized model (set up with parameter that leads to a weaker magnetic field collimation compared to the reference case). Table \ref{8} shows the parameter of the optimized and non optimized models.

\begin{table}[h]
\centering

\begin{tabular}{c}
Optimized Model  \\
\end{tabular}

\begin{tabular}{c | c | c | c}
\hline
$\Gamma$ & $L$ (m) & $D$ (m) & $\vec{B}$ ($10^{-3}$ T) \\
\hline
$0.75$ & $0.9$ & $1.75$ & ($0.9$,$0.3$,$0.3$) \\
\hline
\end{tabular}

\begin{tabular}{c}
Non optimized Model  \\
\end{tabular}

\begin{tabular}{c | c | c | c}
\hline
$\Gamma$ & $L$ (m) & D (m) & $\vec{B}$ ($10^{-3}$ T) \\
\hline
$0.9$ & $1.25$ & $2.5$ & ($1$,$0$,$0$) \\
\hline
\end{tabular}

\caption{Parameters of the optimized and non optimized high resolution models}
\label{8}
\end{table}

The evolution of the kinetic (Fig 14A), current (Fig 14B) and total helicity (Fig 14D) show larger averaged values in the optimized model, as well as an averaged magnetic energy almost 20 times larger compared to the non optimized model, pointing out that the optimization trends obtained by the parametric study leads to an enhancement of the dynamo loop even if the system evolution is non stationary.

\begin{figure}[h]
\includegraphics[width=0.5\columnwidth]{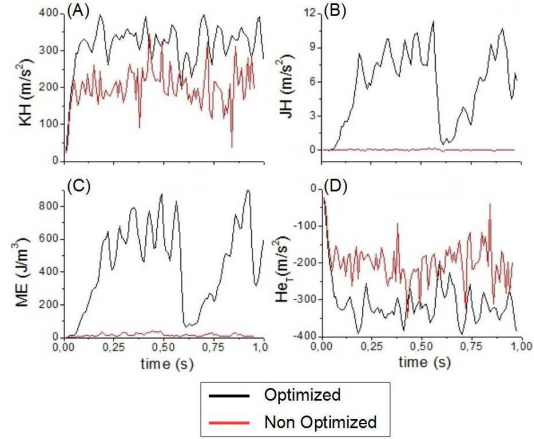} 
\caption{Kinetic helicity(A), current helicity (B), magnetic energy (C) and total helicity (D) in the high resolution simulations for the optimized (black line) and non optimized (red line) models.}
\label{14}
\end{figure}

The helical flows (green lines) in the optimized model (Fig 15B) indicate a focused whirl with the local maximum of the magnetic field (purple iso-contour $|B| = 0.1$ T) located along the whirl vortex, leading to an efficient collimation of the magnetic field lines (black lines) by the impinging flow. For the non optimized model (Fig 15A) the whirl is wider and the local maximum of the magnetic field is located around the whirl (purple iso-countour $|B| = 0.02$ T), pointing out that the magnetic field lines collimation is less efficient compared to the optimized model.

\begin{figure}[h]
\includegraphics[width=0.5\columnwidth]{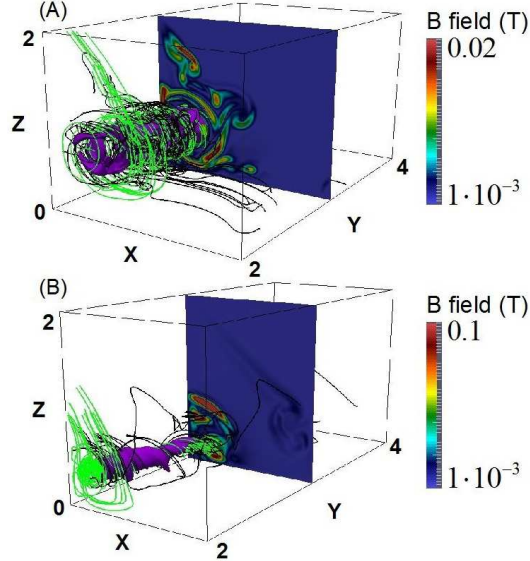} 
\caption{(Color online) Magnetic (black) and velocity (green) field stream lines for the non optimized (A) and optimized cases. Isocontour (pink) of the magnetic field is $0.02$ T ($0.1$ T) for the non optimized (optimized) model. Magnetic field module is plotted at Y = 2 plane (color bar).}
\label{15}
\end{figure}

The non optimized model has a region of anti-alignment between velocity and magnetic fields near the whirl vortex (Fig 16A), correlated with a local minimum of the magnetic field (white arrow). In addition, the local maximum of the magnetic field is correlated with a region of velocity and magnetic fields alignment around the whirl (black arrow). The whirl in the optimized model (Fig 16B) is more focused and no anti-alignment region is observed nearby the vortex. It should be noted that the whirl in the optimized model is located closer to the impeller, particularly to the base of the impeller, leading to a stronger effect of the ferromagnetic boundary conditions on the whirl evolution and an enhancement of the dynamo loop, effect already observed in previous studies \citep{Varela2017}.

\begin{figure}[h]
\includegraphics[width=0.5\columnwidth]{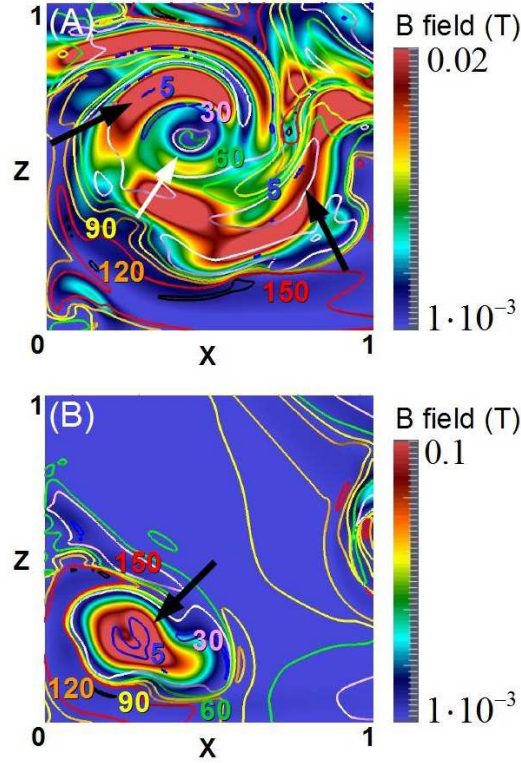} 
\caption{(Color online) Module of the magnetic field (color bar) and angle between velocity and magnetic fields (color isocontours) for the non optimized (A) and nonn optimized (B) models. The region of parallel alignment between the velocity and the magnetic fields is indicated by a black arrow and the region of anti-alignment by a white arrow.}
\label{16}
\end{figure}

In summary, the optimization trends obtained in the parametric studies are validated in a model with non steady evolution and higher turbulence degree, leading to an improvement of the magnetic field collimation and an enhancement of the magnetic energy of the system.

\section{Discussion \label{sec:discussion}}

The present study identifies several optimization trends to enhance the magnetic field collimation by helicoidal flows in between the impeller blades for a numerical set up inspired in the VKS experiment. Among the models analyzed in the parametric study, including different parameter set ups changing the remnant magnetic field orientation and intensity, impinging velocity field due to Ekman pumping and the impeller dimensions, several configurations show the generation of stronger magnetic fields compared with the reference case, namely the TM73 impeller configuration rotating in the unscooping direction.

The magnetic field lines collimation is more efficient if the remnant magnetic field is oriented in the radial direction followed by the vertical orientation, although the azimuthal orientation, dominant in the experiment, is the less efficient. The device optimization requires a remnant magnetic field not purely oriented in the azimuthal direction, because if there is a non negligible poloidal term (vertical + radial), the magnetic field is more efficiently collimated and the magnetic energy of the system increases. In addition, increasing the intensity of the remnant magnetic field also leads to a larger content of the system magnetic energy, although the enhancement is smaller as the remnant magnetic field intensity rises, pointing out that the improvement of the VKS experiment performance using this procedure is constrained.

A larger impinging velocity by Ekman pumping, increasing $\Gamma$ from $0.8$ to $0.9$, leads to a drop of the magnetic field collimation efficiency, however the collimation slightly improves if the Ekman pumping is reduced from $\Gamma = 0.8$ to $0.7$. Consequently, the optimal configuration requires a $\Gamma$ between $0.7$ and $0.8$. 

If the impeller geometry is modified, the largest enhancement of the magnetic field collimation is driven in configurations with a ratio blade height / base length  in between $[0.375-0.5]$ for an azimuthal RMF orientation, and $0.5$ for radial and vertical orientations. This model is an idealized version of the real VKS experiment, the blades are not straight neither parallel, although the analysis conclusions are valid in the impeller regions that satisfy this ratio, leading to an enhanced magnetic field collimation. Nevertheless, the conclusion is only suitable for configurations with comparable Ekman pumping than the reference case, so a dedicated parametric study is required to identify the optimal $\Gamma$ if the impeller geometry changes.  

The implementation of present study optimization trends in VKS experiment is straightforward for the impeller dimensions or the impinging flows, because these improvements can be achieved by the modification of the blade curvature (topology) and the ration blade height / base length. On the other hand, the effect of the RMF on VKS performance is more subtle, because it is linked to the preconditioning of the impeller before the experiment. The RMF in the impeller depends on the magnetic fields driven by dynamo action and external sources in previous VKS runs, so it is difficult to control the RMF orientation and intensity. Due to the properties of the magnetic field driven by dynamo action on VKS, the RMF orientation is mainly toroidal so it cant be consider as an experimental parameter to be modified in present VKS configuration, although in future VKS experiments or other liquid metal experiments, the RMF can be modified inducing currents in the impeller. Consequently, even if the conclusions derived from RMF orientation analysis cant be applied in present VKS experiment, we consider this study a relevant theoretical exercise for future liquid metal experiment designs. Beside that, the optimization trends related to the RMF intensity can be tested in VKS, because the intensity of the RMF is determined by the strength of the magnetic field driven by dynamo action and external sources, so it can be considered as an experimental parameter even though it is difficult to control. It should be noted that we don't include in the analysis feedback effects between different experimental parameters, although the trends found in the study are robust because we identify VKS performance optimization as an improvement of the magnetic field collimation by the flows, further confirmed by an enhancement of the dynamo loops, so that tendency must remain valid in first approximation.

The analysis of the dynamo loop features using the mean field dynamo theory shows that the $\Omega$-$\alpha$ dynamo loop is between one and two orders of magnitude larger compared to the $\alpha^{2}$ dynamo loop in the simulations, although the hypothetical $\alpha^{2}$ dynamo loop is enhanced if the RMF is oriented in the vertical or radial direction, the ratio between the impeller blade height and the impeller base length is $0.5$ or the Ekman pumping drops to $\Gamma = 0.7$. Increasing the RMF intensity leads to a weaker $\alpha^{2}$ dynamo loop, except for the model with the RMF oriented in the radial direction. The models discussed in the dynamo loop analysis are steady state simulations because the turbulence level is not large enough to drive the oscillatory evolution observed in models with $R_{e} = 1000$, so the hypothetical $\Omega$-$\alpha$ dynamo loop is dominant compared with the $\alpha^{2}$ dynamo loop, although the configuration with the RMF oriented in the radial or vertical directions have a large contribution of the $\alpha^{2}$ dynamo loop. We also performed simulations with larger resolution and non steady evolution of the model to confirm the optimization trends observed in the dynamo loop analysis of steady state models. The optimized high resolution model shows an enhancement of the dynamo loop and a more efficient collimation of the magnetic field lines compared to the non optimized model, validating the optimization trends.

In the real VKS operation regime, $R_{e} \approx 10^{6}$ so the turbulence level is much larger than in the simulations and the $\alpha^{2}$ dynamo loop must be further reinforced. On the other hand, the enhancement of the differential rotation via e.g. differential rotation of the impeller may counterbalance this effect, and favor local $\alpha- \Omega$ or $\alpha^{2}-\Omega$ dynamo mechanisms. Several of the configurations analyzed show an important enhancement of the $\alpha^{2}$ dynamo loop even for steady state model, pointing out that such configurations in VKS operation regimes could show a dominant $\alpha^{2}$ rather than $\Omega\alpha^{2}$ or $\Omega\alpha$ dynamo loops. This result opens the possibility to operate the VKS device using configurations optimized to generate dynamos with dominant $\alpha^{2}$ or $\Omega\alpha^{2}$ dynamo loops.

Next table \ref{9} summarizes the study results indicating the optimal VKS configuration in the parameter space analyzed to enhance the magnetic field collimation:

\begin{table}[h]
\centering
\begin{tabular}{| c | c | c | c | c |}
\hline
$L/D$ & $\Gamma$ & $|\vec{B}|$ \\
\hline
$0.375 - 0.5$ & $0.7 - 0.8$ & $ME \approx B_{0}^{1.68}$ \\
\hline
\end{tabular}

\caption{Optimal configuration.}
\label{9}
\end{table}

\begin{acknowledgments}
We have also received funding by the Labex PALM/P2IO/ LaSIPS (VKStars grant number 2013-02711), INSU/ PNST and ERC PoC grant 640997 Solar Predict. We thank S. Brun, B. Dubrulle, C. Nore and the VKS team for fruitful discussions. \\
{\bf Disclaimer:} This manuscript has been authored by UT-Battelle, LLC under Contract No. DE-AC05- 00OR22725 with the U.S. Department of Energy. The United States Government retains and the publisher, by accepting the article for publication, acknowledges that the United States Government retains a non-exclusive, paid-up, irrevocable, world-wide license to publish or reproduce the published form of this manuscript, or allow others to do so, for United States Government purposes. The Department of Energy will provide public access to these results of federally sponsored research in accordance with the DOE Public Access Plan (http://energy.gov/downloads/doe-public-access-plan).
\end{acknowledgments}

\appendix
\section{Model summary}
\label{sec:Model_summary}

Table \ref{10} shows the model names, the remnant magnetic field orientation and intensities, $\Gamma$ values, impeller base lengths $D$ and blade heights $L$. The reference model is the case with azimuthal RMF orientation of intensity $10^{-3}$ T, $\Gamma = 0.8$, $L = 1.0$ m and $D = 2$ m, identified as the TM 73 impeller configuration rotating in the unscooping direction. The name code shows the remnant magnetic field orientation ("BX", "BY" or "BZ") and the subindex indicates the parameter modified regarding the reference model (except models $BY_{B0}$ and $BZ_{B0}$ where only the RMF orientation is different compared to the reference case).

\begin{table}[h]
\centering

\begin{tabular}{c | c c c c}
Model & $\vec{B}$ ($10^{-3}$ T) & $\Gamma$ & $L$ (m) & $D$ (m) \\
\hline
TM 73 & (1,0,0) & 0.8 & 1 & 2 \\
$BX_{B0/2}$ & (0.5,0,0) & 0.8 & 1 & 2 \\
$BX_{5B0}$ & (5,0,0) & 0.8 & 1 & 2 \\
$BX_{10B0}$ & (10,0,0) & 0.8 & 1 & 2 \\
$BX_{\Gamma=0.7}$ & (1,0,0) & 0.7 & 1 & 2 \\
$BX_{\Gamma=0.9}$ & (1,0,0) & 0.9 & 1 & 2 \\
$BX_{L=0.75}$ & (1,0,0) & 0.8 & 0.75 & 2 \\
$BX_{L=1.25}$ & (1,0,0) & 0.8 & 1.25 & 2 \\
$BX_{D=1.0}$ & (1,0,0) & 0.8 & 1 & 1.0 \\
$BX_{D=1.5}$ & (1,0,0) & 0.8 & 1 & 1.5 \\
$BX_{D=2.5}$ & (1,0,0) & 0.8 & 1 & 2.5 \\
$BY_{B0}$ & (0,1,0) & 0.8 & 1 & 2 \\
$BY_{B0/2}$ & (0,0.5,0) & 0.8 & 1 & 2 \\
$BY_{5B0}$ & (0,5,0) & 0.8 & 1 & 2 \\
$BY_{10B0}$ & (0,10,0) & 0.8 & 1 & 2 \\
$BY_{\Gamma=0.7}$ & (0,1,0) & 0.7 & 1 & 2 \\
$BY_{\Gamma=0.9}$ & (0,1,0) & 0.9 & 1 & 2 \\
$BY_{L=0.75}$ & (0,1,0) & 0.8 & 0.75 & 2 \\
$BY_{L=1.25}$ & (0,1,0) & 0.8 & 1.25 & 2 \\
$BY_{D=1.0}$ & (0,1,0) & 0.8 & 1 & 1.0 \\
$BY_{D=1.5}$ & (0,1,0) & 0.8 & 1 & 1.5 \\
$BY_{D=2.5}$ & (0,1,0) & 0.8 & 1 & 2.5 \\
$BZ_{B0}$ & (0,0,1) & 0.8 & 1 & 2 \\
$BZ_{B0/2}$ & (0,0,0.5) & 0.8 & 1 & 2 \\
$BZ_{5B0}$ & (0,0,5) & 0.8 & 1 & 2 \\
$BZ_{10B0}$ & (0,0,10) & 0.8 & 1 & 2 \\
$BZ_{\Gamma=0.7}$ & (0,0,1) & 0.7 & 1 & 2 \\
$BZ_{\Gamma=0.9}$ & (0,0,1) & 0.9 & 1 & 2 \\
$BZ_{L=0.75}$ & (0,0,1) & 0.8 & 0.75 & 2 \\
$BZ_{L=1.25}$ & (0,0,1) & 0.8 & 1.25 & 2 \\
$BZ_{D=1.0}$ & (0,0,1) & 0.8 & 1 & 1.0 \\
$BZ_{D=1.5}$ & (0,0,1) & 0.8 & 1 & 1.5 \\
$BZ_{D=2.5}$ & (0,0,1) & 0.8 & 1 & 2.5 \\
\hline
\end{tabular}

\caption{Model summary}
\label{10}
\end{table}

\section{Autocorrelation}
\label{sec:Autocorrelation}
Definition of the time autocorrelation function of the velocity averaged in the azimuthal/toroidal direction ($F(\tau)$):
$$ F_{i}(\tau) =  \frac{\int_{t_{0}}^{t_{f}} \left\langle u_{i}(t) \right\rangle \left\langle u_{i}(t + \tau) \right\rangle dt}{\int_{t_{0}}^{t_{f}} \left\langle u_{i}(t) \right\rangle^2} $$
with $\langle \rangle$ indicating an average in the toroidal direction. The autocorrelation time of the velocity averaged in the toroidal direction ($C_{\tau}$) is defined as the time (t) when $F(t = t_{0} + \tau)<F(t_{0})/2$, with $i=1,2,3$ the velocity components and $\tau$ the time lag.

Definition of the length autocorrelation function of the velocity averaged in the toroidal direction ($F(d)$):
$$ F_{i}(d) =  \frac{\int_{r_{0}}^{r_{f}} \left\langle u_{i}(r) \right\rangle \left\langle u_{i}(r + d) \right\rangle dr}{\int_{r_{0}}^{r_{f}} \left\langle u_{i}\right\rangle (r)^2} $$
the autocorrelation length of the velocity averaged in the toroidal direction ($C_{d}$) is defined as the length (r) where $F(r = r_{0} + d)<F(r_{0})/2$ with $d$ the length lag.

Table~\ref{11} shows the autocorrelation factor for each model:

\begin{table}[h]
\centering

\begin{tabular}{c | c }
Model & $C_{d}$ / $C_{\tau}$ (m/s) \\
\hline
Reference & 0.25 \\
$BY_{B0}$ & 0.19 \\
$BZ_{B0}$ & 0.18 \\
$BX_{B0/2}$ & 0.17 \\
$BX_{5B0}$ & 0.18 \\
$BX_{10B0}$ & 0.18 \\
$BY_{10B0}$ & 0.19 \\
$BZ_{10B0}$ & 0.19 \\
$\Gamma_{0.7}$ & 0.18 \\
$\Gamma_{0.9}$ & 0.20 \\
$L_{0.75}$ & 0.17 \\
$L_{1.25}$ & 0.27 \\
$D_{1.5}$ & 0.23 \\
$D_{2.5}$ & 0.18 \\
\hline
\end{tabular}

\caption{Autocorrelation factor.}
\label{11}
\end{table}

\bibliographystyle{plainnat}

\begin{thebibliography}{100}

\bibitem{2006GApFD.100..281N} Nataf, H. et al {\it Geophys. Astro. Fluid}, {\bf 100}, 281, (2006).        
\bibitem{ZAMM:ZAMM19840640913} Moffatt, H. K. et al {\it ZAMM}, {\bf 64}, 1521, (1984).  
\bibitem{2004ApJ...614.1073B} Brun, A. S. et al {\it ApJ}, {\bf 614}, 1073, (2004).  
\bibitem{1.1331315} Stieglitz, R. et al {\it Phys. Fluids}, {\bf 13}, 561, (2001).  
\bibitem{2000PhRvL..84.4365G} Gailitis, A et al {\it Phys. Rev. Lett.}, {\bf 84}, 4365, (2000).  
\bibitem{2007PhRvL..98d4502M} Monchaux, R. et al {\it Phys. Rev. Lett.}, {\bf 98}, 044502, (2007).  
\bibitem{03091920701523410} Pétrélis, F. et al {\it Geophys. Astro. Fluid}, {\bf 101}, 289, (2007).  
\bibitem{2009EL.....8739002G} Gissinger, C. et al {\it EPL}, {\bf 87}, 39002, (2009).  
\bibitem{PhysRevLett.101.104501} Laguerre, R. et al {\it Phys. Rev. Lett.}, {\bf 101}, 104501, (2008).  
\bibitem{PhysRevLett.109.024503} Ravelet, F. et al {\it Phys. Rev. Lett.}, {\bf 109}, 024503, (2012).  
\bibitem{PhysRevLett.104.044503} Giesecke, A. et al {\it Phys. Rev. Lett.}, {\bf 104}, 044503, (2010).  
\bibitem{PhysRevE.91.013008} Nore, C. et al {\it Phys. Rev. Lett. E}, {\bf 91}, 013008, (2015).  
\bibitem{Varela2017} Varela, J. et al {\it Phys. Plasma}, {\bf 24}, 053518, (2017).  
\bibitem{PhysRevE.92.063015} Varela, J. et al {\it Phys. Rev. E}, {\bf 92}, 063015, (2015).  
\bibitem{PhysRevE.88.013002} Miralles, S. et al {\it Phys. Rev. E}, {\bf 88}, 013002, (2013).  
\bibitem{2007ApJS..170..228M} Mignone, A. et al {\it ApJs}, {\bf 170}, 228, (2007).  
\bibitem{1367-2630-14-1-013044} Boisson, J. et al {\it New J. Phys.}, {\bf 14}, 013044, (2012).  
\bibitem{Nore} Nore, C. et al {\it Europhys. Lett.}, {\bf 114}, 65002, (2016).  
\bibitem{2014NJPh...16h3001F} Faranda, D. et al {\it New J. Phys.}, {\bf 16}, 083001, (2014).  
\bibitem{Ravelet} Ravelet, F.l {\it PhD Thesis, Ecole Polytechnique} (2005).  
\bibitem{Moffatt78} Moffatt, H. et al {\it Magnetic Field Generation in Electrically Conducting Fluids. Cambridge et al., Cambridge University Press} (1983).  
\bibitem{Ossendrijver2003} Ossendrijver, M. et al {\it Astron. Astrophys. Rev.}, {\bf 11}, 287, (2003).  
\bibitem{Beck1996} Beck, R. et al {\it Annu. Rev. Astron. Astrophys}, {\bf 34}, 115, (1996).  
\bibitem{Aubert2015} Aubert, J. et al {\it Geophys J Int}, {\bf 203}, 1738, (2015).  
\bibitem{1980AN....301..101R} R{\"a}edler, K. et al {\it Astron. Nachr.}, {\bf 301}, 101, (1980).  
\bibitem{Durbin} Durbin, P. et al {\it Statistical Theory and Modeling for Turbulent Flows, Second Edition, John Wiley $\&$ Sons, Ltd} (2010).  
\bibitem{Ott} Ott, E. et al {\it Phys. of Plasmas}, {\bf 5}, 1636, (1998).  
\bibitem{Simitev} Simitev, R. et al {\it ApJ}, {\bf 810}, 80, (2015).  
\bibitem{Yadav} Yadav, R. et al {\it ApJL}, {\bf 833}, L28, (2016).  
\bibitem{Brown} Brown, B. et al {\it ApJ}, {\bf 731}, 69, (2011).  
\bibitem{Brown2} Brown, B. et al {\it ApJ}, {\bf 711}, 424, (2010).  
\bibitem{Dormy} Dormy, E. et al {\it Proc Int Astron Union}, {\bf 8}, 163, (2012).  
\bibitem{Chabrier} Chabrier, G. et al {\it A\&A}, {\bf 446}, 1027, (2006). 
\bibitem{Schubert} Schubert, G. et al {\it ApJ}, {\bf 532}, L149, (2000). 
\bibitem{Gissinger} Gissinger, C. et al {\it EPL}, {\bf 82}, 29001, (2008). 

\end{thebibliography}

\end{document}